\documentclass[aip,apl,graphicx, 12pt]{revtex4}
\usepackage{graphicx}
\draft 

\begin{document}

\title{Pulsed laser deposition of SrTiO$_{3}$/LaGaO$_{3}$ and SrTiO$_{3}$/LaAlO$_{3}$: plasma plume effects}

\author{C. Aruta}
\email{aruta@na.infn.it} \affiliation{CNR-SPIN and Dipartimento di
Scienze Fisiche, Complesso Universitario di Monte Sant'Angelo, Via
Cintia, I-80125 Napoli, Italy}
\author{S. Amoruso}
\affiliation{CNR-SPIN and Dipartimento di Scienze Fisiche,
Complesso Universitario di Monte Sant'Angelo, Via Cintia, I-80125
Napoli, Italy}
\author{R. Bruzzese}
\affiliation{CNR-SPIN and Dipartimento di Scienze Fisiche,
Complesso Universitario di Monte Sant'Angelo, Via Cintia, I-80125
Napoli, Italy}
\author{X. Wang}
\affiliation{CNR-SPIN and Dipartimento di Scienze Fisiche,
Complesso Universitario di Monte Sant'Angelo, Via Cintia, I-80125
Napoli, Italy}
\author{D. Maccariello}
\affiliation{CNR-SPIN and Dipartimento di Scienze Fisiche,
Complesso Universitario di Monte Sant'Angelo, Via Cintia, I-80125
Napoli, Italy}
\author{F. Miletto Granozio}
\affiliation{CNR-SPIN and Dipartimento di Scienze Fisiche,
Complesso Universitario di Monte Sant'Angelo, Via Cintia, I-80125
Napoli, Italy}
\author{U. Scotti di Uccio}
\affiliation{CNR-SPIN and Dipartimento di Scienze Fisiche,
Complesso Universitario di Monte Sant'Angelo, Via Cintia, I-80125
Napoli, Italy}

\date{\today}

\begin{abstract}

Pulsed laser deposition of SrTiO$_{3}$/LaGaO$_{3}$ and
SrTiO$_{3}$/LaAlO$_{3}$ interfaces has been analyzed with a focus
on the kinetic energy of the ablated species. LaGaO$_{3}$ and
LaAlO$_{3}$ plasma plumes were studied by fast photography and
space-resolved optical emission spectroscopy. Reflection high
energy electron diffraction was performed proving a layer-by-layer
growth up to $10^{-1}$ mbar oxygen pressure. The role of the
energetic plasma plume on the two-dimensional growth and the
presence of interfacial defects at different oxygen growth
pressure has been discussed in view of the conducting properties
developing at such polar/non-polar interfaces.
\end{abstract}


\maketitle

The recent discovery of a 2-dimensional electron gas (2DEG) at the
interface between two insulators, e.g. the polar LaAlO$_{3}$ (LAO)
 and the non-polar SrTiO$_{3}$ (STO), raised great interest for both fundamental and applicative perspectives.
 The well known interpretation in terms of the so-called polar catastrophe has certainly some significance,
 but other mechanisms may as well contribute to the observed phenomenology. Since pulsed laser deposition (PLD)
 is the most used technique for the growth of these interfaces, an analysis of the growth process can be helpful in trying
 to better elucidate features related to the 2DEG formation and to pose some constraints to the various physical mechanisms
 involved. While the polar catastrophe scenario applies to perfect
interfaces, alternative mechanisms are based on the possible
doping role of point defects. A first mechanisms is related to the
creation of oxygen vacancies in the STO crystal used as substrate
during PLD. Oxygen vacancies are known to result in an electron
doping, leading to bulk conductivity, and even superconductivity
below 400 mK. The possible role of such defects in the 2DEG
formation at the SrTiO$_{3}$/LaAlO$_{3}$ (STO/LAO) interface was
already addressed in the seminal paper of Ohtomo and
Hwang\cite{Ohtomo_Hwang}. Several following papers
\cite{KalabukhovPRB2007, HuijbenAdvMat, SiemonsPRL2007} discussed
the dependence of the electrical transport properties on the
oxygen pressure during PLD of the LAO overlayer, typically in the
range 10$^{-6}$ and 10$^{-3}$ mbar. Kalabukhov et al.
\cite{KalabukhovPRB2007} demonstrated that at 10$^{-6}$ mbar a
large number of dislocations and of oxygen vacancies is generated.
It was also suggested that this may be connected with the impact
of high kinetic energy species. This leads to the second possible
source of point defects: if energetic La ions hit the STO surface,
some cation intermixing may also take place, resulting in a
chemical doping of the STO surface with potential effect on the
2DEG formation \cite{HuijbenAdvMat}. It is tempting to increase
the oxygen pressure during LAO growth, in order to slow down the
impinging species; however, this modifies the growth mode. Huijben
et al. found a crossover from 2D layer-by-layer to the island
growth mode at 10$^{-2}$ mbar \cite{HuijbenAdvMat}. Maurice et al.
\cite{Maurice} also showed that above 10$^{-1}$ mbar STO/LAO had a
rough surface and poor structure. In spite of the relevance of the
issues, an analysis of the PLD plume and an exact determination of
the species kinetic energies during the fabrication of STO/LAO
interfaces is still lacking.\\
In this letter we report a study of
the PLD process of SrTiO$_{3}$/LaGaO$_{3}$ (STO/LGO) and STO/LAO,
carried out simultaneously with high energy electron diffraction
(RHEED) on the growing samples, at different oxygen background
pressures. 2DEG formation in STO/LGO was very recently reported
\cite{Perna_APL}, and the analysis of both systems could help
elucidating possible differences related to their specific
composition.\\

Films of LGO\cite{Perna_APL} and LAO\cite{Savoia} were deposited
on single TiO$_{2}$-terminated STO substrates. The ablation was
carried out by irradiating LGO or LAO rotating targets with laser
pulses of 25 ns duration (full width half maximum) delivered by a
KrF excimer laser at a repetition rate of 1 Hz at different oxygen
pressures in the range from high vacuum (HV) up to 10$^{-1}$ mbar.
The substrate was positioned at a distance of 3.5 cm from the
target surface and its temperature was fixed at 800$^{\circ}$C.
This procedure enables the achievement of a conducting 2DEG in
both STO/LGO and STO/LAO, provided the film thickness reaches the
4 unit cells (u.c.) threshold. The full electrical transport
characterization, together with a detailed analysis of the
microstructural properties of the interfaces, is reported in
ref.\cite{Perna_APL}. Conducting interfaces are routinely achieved
in the samples grown in the range 10$^{-4}$ - 10$^{-2}$ mbar, with
no clear dependence of the transport properties on the oxygen
pressure. During the growth process, the plasma plume was
investigated by fast photography and space-resolved optical
emission spectroscopy,
following the plume expansion from the target to the substrate and identifying plume composition \cite{AmorusoJAP2010}.\\
RHEED analysis was performed during the growth. Intensity
oscillations were observed up to 10$^{-1}$ mbar, as shown in Figs.
1 (a) and (c). However, while the oscillations are not regular at
such a high pressure, very good oscillations are obtained at lower
oxygen pressure \cite{Perna_APL}. The oscillations indicate the
two-dimensional (2D) layer-by-layer growth mode. In both LGO and
LAO, the initial intensity drop is partially related to an
extrinsic factor, i.e. the slight variation of the optimal
diffraction conditions with respect to the STO substrate. In the
case of LGO, an increase of the RHEED background is superimposed
on the oscillations for deposition time between about 50 and 150
s. This is due to the progressive formation of the LGO layers
characterized by a larger scattering efficiency. After completing
the deposition, the RHEED signal increases (not shown here), as
expected when an ordering process takes place at the surface. Our
key result, in view of the following discussion, was the
capability to achieve flat LGO and LAO surfaces even at 10$^{-1}$
mbar, as qualified by the streaky RHEED patterns shown in Figs.
1(b, d), recorded at the end of the growth of 12 u.c. thick films.
This pushes to a much higher pressure the limit for the 2D growth,
previously reported as 10$^{-3}$ mbar \cite{HuijbenAdvMat}.
However, the samples grown at
10$^{-1}$ mbar are insulating.\\

Fig. 2(a) reports typical images of the LGO plume emission at
three different delays after the laser pulse, while spatially
resolved emission spectra (in the range 370 - 630nm), collected at
1.6 $\mu$s delay at 10$^{-3}$ - 10$^{-1}$ mbar, are shown as an
example in Fig. 2(b). The spectra show intense emissions from both
La and Ga (Al in the case of LAO) neutrals, and from the Lanthanum
oxide LaO. Qualitatively similar pictures are obtained for the LAO
plume. The image at low pressure shows that the region of maximum
emission is located at the rear part of the plume (free expansion
regime), while it is shifted toward the front at increasing
pressure (shock wave regime), similarly to what has been observed
earlier \cite{AmorusoJAP2010}. Sequences of images as in Fig. 2
allow one to follow the plume propagation and henceforth to
determine the maximum kinetic energy of the various species at the
substrate position (see Table I). The data show that a crossover
from free to braked expansion takes place at about 10$^{-2}$ mbar,
as indicated by the progressive changes of the plume shape in Fig.
2(a). In particular, at 10$^{-2}$ mbar the interaction with the
background gas starts promoting plume excitation and oxidation of
the ablated species at the plume front, while only slightly
influencing the maximum kinetic energy of the species finally
impacting the substrate. By further increasing the oxygen pressure
to 10$^{-1}$ mbar, the plume is more braked and the kinetic energy
of the impinging particles is strongly reduced. Moreover, as a
consequence of the interaction with the background oxygen,
emission from LaO at the plume front appears at 10$^{-2}$ mbar,
and becomes significantly enhanced at the larger pressure of
10$^{-1}$ mbar.
On the contrary, neither Ga nor Al oxides are observed.\\

On the basis of the data, we can now comment on the main issues of
the paper. First we discuss La/Sr intermixing at the STO surface
possibly induced by the impact of the energetic species. The
ablated species lose their kinetic energy when impinging on the
growing surface by elastic and inelastic collisions. The typical
loss rate for this process is of $\approx$ 100 eV/nm
\cite{Anders2002}. Then, the maximum kinetic energies of Table I
fix a limit to the maximum subplantation depth, thus to the
possible intermixing region close to the interface. Accordingly,
below 10$^{-2}$ mbar oxygen pressure, the maximum La-subplantation
depth is less than 0.6 nm (about 1.5 u.c.), while at 10$^{-1}$
mbar the kinetic energy is too low to give any La-subplantation.
Considering that below 10$^{-2}$ mbar the kinetic energy
distribution within the ablation plume (not shown) indicates a
most probable kinetic energy of $\approx$ 10 eV, which rapidly
decreases at larger kinetic energies values, the most energetic
species constitute only a minor fraction of the plume. We
therefore conclude that at usual growth conditions the impact
dynamics may provide a low or null surface cation intermixing.
These findings are compatible with the high-resolution scanning
transmission electron microscopy images and electron energy loss
spectroscopy profiles across the STO/LGO and STO/LAO interfaces
reported in ref. \cite{Perna_APL}, showing interfaces
that are sharp at the limit of instrumental resolution.\\

As for the role of the oxygen vacancies, two issues are in order:
(a) the impact of energetic species creating vacancies through
oxygen sputtering at the STO surface, and (b) deposition of oxygen
deficient LGO (LAO) overlayer which can get the lacking oxygen at
expenses of the STO substrate through diffusion
\cite{Uedono_JAP2002, LippertICPEPA7}. As for  mechanism (a), a
simple evaluation can be based on elastic collisions between the
impinging species and the oxygen atoms at the STO surface. By
resting on tabulated bond strengths of Ti-O and Sr-O \cite{CRC},
the energy required to break the oxygen bonds and kick it off the
surface is $\approx$ 10 eV. This simple estimate is consistent
with the values obtained by first principles methods
\cite{Carrasco_2005}. The maximum kinetic energy of the impinging
atoms of Table I can be therefore sufficient to kick off oxygen
atoms for pressures up to 10$^{-2}$ mbar. At larger pressure, this
process is hindered by the significant reduction of the maximum
kinetic energy. Nevertheless, the over-pressure due to the
presence of the ablation plume at the substrate surface could
further limit the efficacy of such a process. As for mechanism
(b), our data suggest that La is the only cation acting as an
oxygen getter (see Fig. 2(b)), and the direct deposition of
plume's LaO molecules is therefore one channel of incorporation of
oxygen in the film. Below 10$^{-2}$ mbar LaO formation is rather
ineffective, and only at 10$^{-1}$ mbar a significant amount of
LaO at the plume front is formed as a consequence of shock wave
\cite{AmorusoJAP2010} and molecular oxygen dissociation
\cite{Camposeo}. In this respect, we notice that distance-time
plots of the plume front propagation (not shown) confirm that the
plume follows a shock-wave-like propagation behavior close to the
substrate surface at 10$^{-1}$ mbar. This suggests that films
deposited at lower pressure can be oxygen deficient, and could get
the lacking oxygen at expenses of the STO substrate, which instead
is not very likely at 10$^{-1}$ mbar. These observations seem to
correlate well with the variation of the conducting properties of
the STO/LGO and STO/LAO, suggesting a possible role of these
mechanisms on the properties of the interfaces. Nevertheless, it
is worth noting that SrTiO$_{3}$/LaMnO$_{3}$ interfaces grown in
similar conditions, at 10$^{-4}$ - 10$^{-2}$ mbar, show an
insulating character \cite{Perna_APL}. This testifies the
complexity of the 2DEG generation, where specific material issues
may also play an important role.\\

Finally, we want to comment on the effects of the particles
kinetic energy on the growth dynamics. As discussed in Ref.
\cite{Willmott2006} kinetic energy can favor 2D growth through
island breakup mechanism with prompt insertion, at very low
coverage, and enhanced surface diffusion, above 50$\%$ monolayer
coverage. Following the approach of ref.\cite{Willmott2006}, the
energy scale of the island break-up mechanism can be estimated by
the binding energy between one island made by two unit cells and
two separated single unit cells. On the basis of the tabulated
bond strengths of La-O, Al-O and Ga-O \cite{CRC}, such energy is
$\approx$ 20 eV in the case of LGO and LAO. When comparing this
figure with the data of Table I, one realizes that the stable 2D
growth is consistent with the growth mechanisms discussed above
only below 10$^{-2}$ mbar. However, while the kinetic energy
strongly drops at the higher pressure, RHEED data still show a 2D
growth (see Fig. 1). We propose that a further amount of energy is
provided by the internal energy of the particles. Actually, the
electronic transitions in the spectra of Fig. 2 are in the 2-3 eV
range. Such values are comparable with the maximum surface
diffusion barrier energy at high coverage reported in
ref.\cite{Willmott2006}; therefore the higher plume internal
energy at 10$^{-1}$ mbar eventually favors surface diffusion
resulting in the 2D growth. The differences in the growth dynamics
at low
and high pressure might finally also reflect on the different conducting properties of the interfaces.\\

In conclusion, we studied PLD of LGO and LAO on STO elucidating
the effects of the background oxygen gas pressure on the ablation
plume and deposited film interfaces. Analysis of oxygen pressure
variation on ablated species kinetic energy and oxidation state
allowed us to provide further insights on some of the mechanisms
(e.g. oxygen vacancy creation, La subplantation, and growth
dynamics) considered to contribute to conductivity of the
interfaces, in addition to the polar catastrophe. Our results also
indicate that 2D growth can be achieved up to 10$^{-1}$ mbar, but
a crossover exists at 10$^{-2}$ mbar since the kinetic energy of
the impinging species changes from tens of eV to tenths of eV as a
consequence of the interaction with the background gas.

\clearpage

\begin{table}
\caption{\label{tab:table1} Maximum kinetic energy (eV) of the
ablated species from LGO and LAO impacting the substrate at
different growth pressures (mbar).}
\begin{ruledtabular}
\begin{tabular}{cccc}
Species & 10$^{-6}$-10$^{-3}$ & 10$^{-2}$ & 10$^{-1}$\\
 \hline
Al & 14$\pm$0.2 & 14$\pm$0.2 & 0.1$\pm$0.02\\
Ga & 35$\pm$0.2 & 35$\pm$0.2 & 0.4$\pm$0.02\\
La & 58$\pm$0.2 & 55$\pm$0.2 & 0.7$\pm$0.02\\
LaO & 52$\pm$0.2 & 52$\pm$0.2 & 0.8$\pm$0.02\\
\end{tabular}
\end{ruledtabular}
\end{table}
\clearpage
\begin{figure}
\includegraphics[width=8 cm]{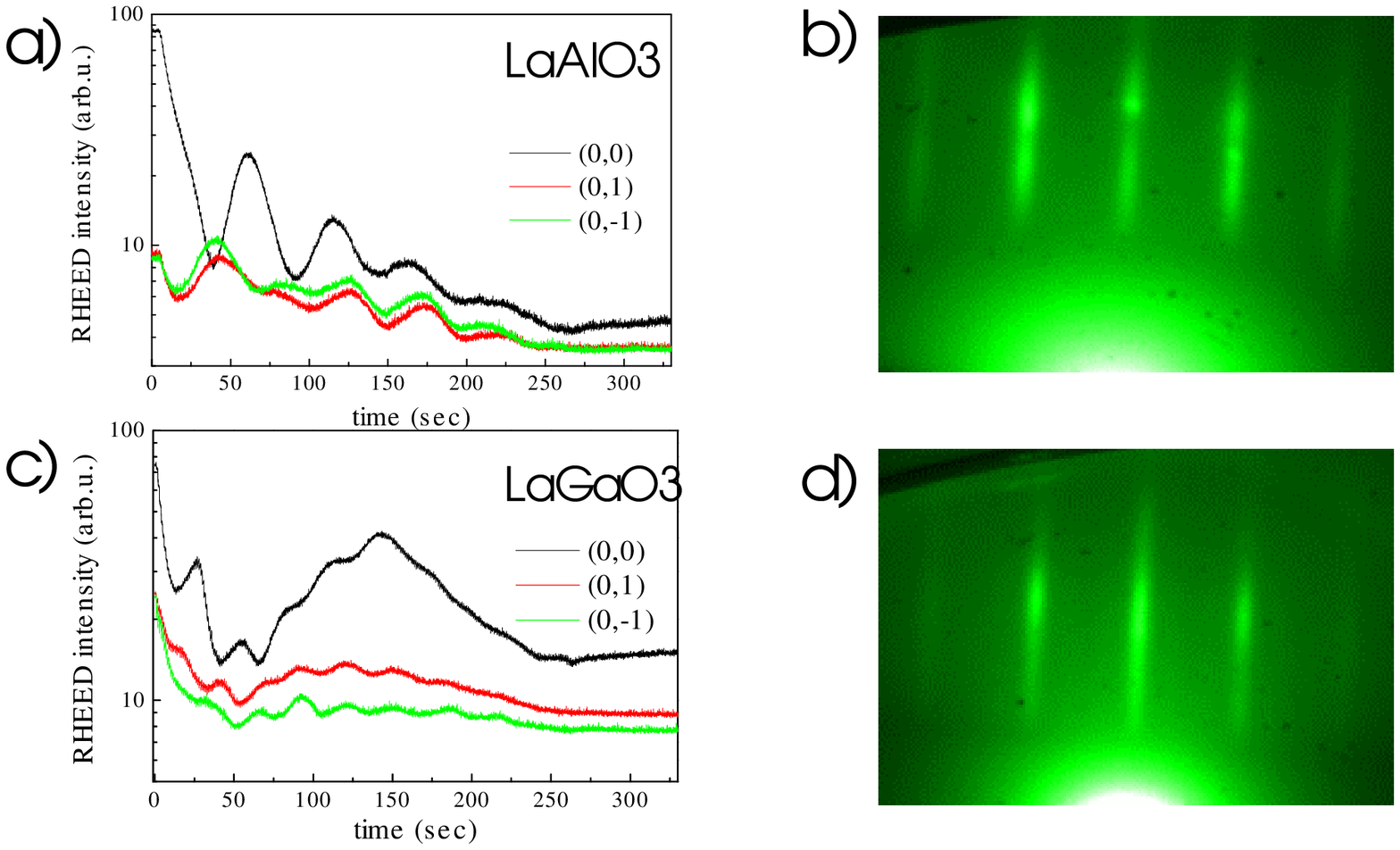}
\caption{(Color online) RHEED oscillations during the initial
phase of (a) LaAlO$_{3}$ and (c) LaGaO$_{3}$ film growth on a
SrTiO$_{3}$ substrate at 10$^{-1}$ mbar of oxygen. RHEED patterns
after the growth of 12 u.c. of (b) LaAlO$_{3}$ and (d)
LaGaO$_{3}$.}
\end{figure}
\clearpage

\begin{figure}
\includegraphics[width=8.5 cm]{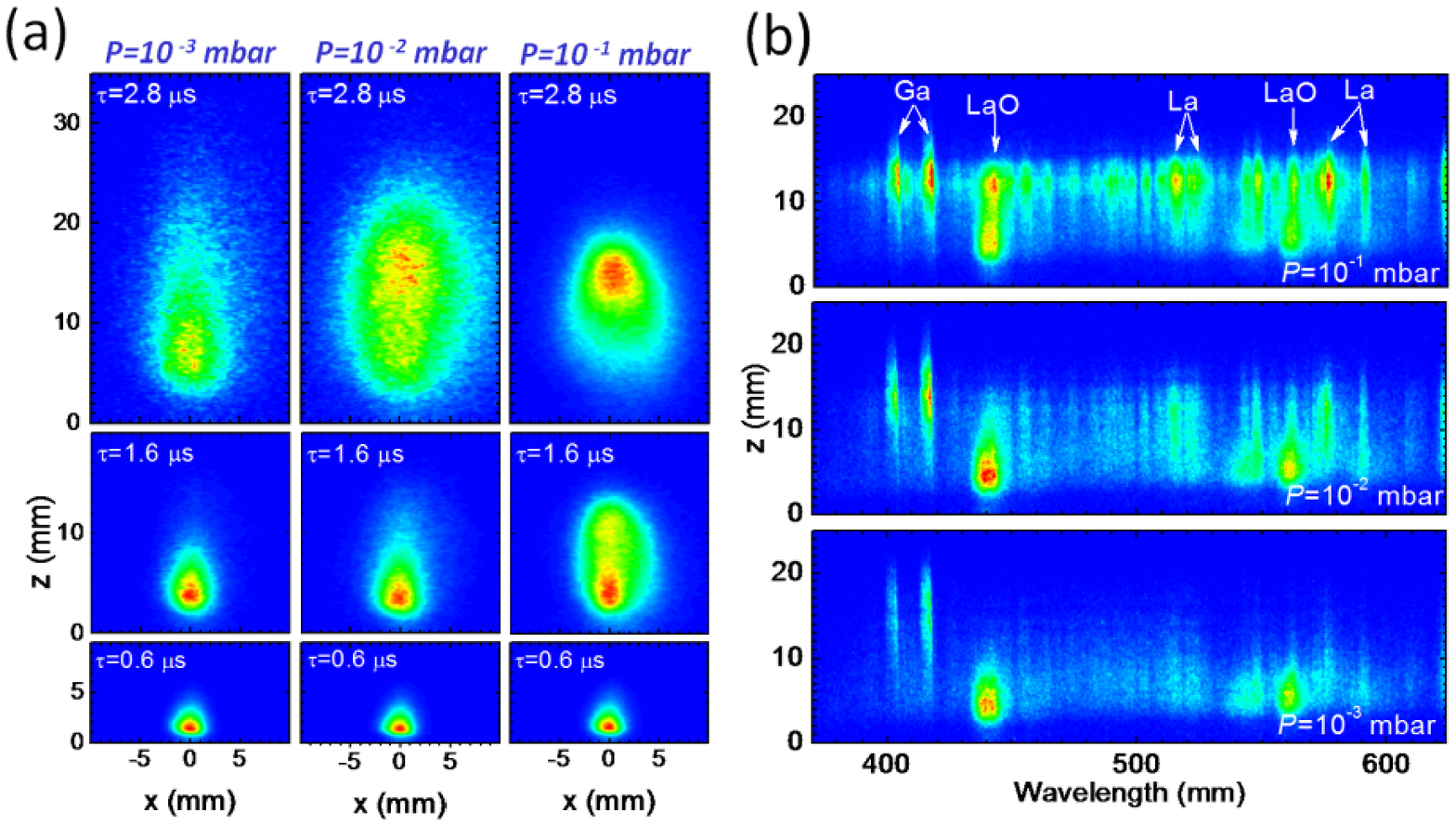}
\caption{(Color online) (a) 2D single-shot images of LGO ablation
plume at three different delays $\tau$ after the laser pulse for
oxygen pressure from 10$^{-3}$ (left) to 10$^{-1}$ mbar (right).
Each image is obtained from a different laser shot and shown in
normalized false color scale. z=0 marks the position of the target
surface. The images at 10$^{-3}$ mbar are also representative of
the plume propagation registered at lower pressure.(b) Emission
spectra of the LGO plume at a delay $\tau$=1.6$\mu$s for three
different oxygen pressures (10$^{-3}$, 10$^{-2}$ and 10$^{-1}$
mbar).}
\end{figure}

\end{document}